# Superconducting Low-beta Nb$_3$Sn Cavity for ATLAS and Future Ion Accelerators

T. Petersen,[1, a)] G. Eremeev,[2] B. Tennis,[2] N. Tagdulang,[2] Y. Zhou,[1] M. Kedzie,[1] B. Guilfoyle,[1] S. Xu,[1, b)] S.V. Kutsaev,[3] R. Agustsson,[3] E. Spranza,[3] P. Davis,[1] G. P. Zinkann,[1] T. Reid,[1] S. Posen,[2] and M. P. Kelly[1]

[1)]*Argonne National Laboratory, Lemont, United States of America*
[2)]*Fermi National Accelerator Laboratory, Batavia, United States of America*
[3)]*RadiaBeam Technologies, Santa Monica, United States of America*

(*Electronic mail: tpetersen@anl.gov)

(Dated: 2 October 2025)

We report on a Nb$_3$Sn-coated low-beta superconducting radio frequency (SRF) cavity intended for accelerating ions. We aim to apply the cavity in ATLAS, our Argonne National Laboratory user facility for nuclear physics studies with ion beams in the energy range of 5-20 MeV/u. The Nb$_3$Sn-coated cavity, a 145 MHz quarter-wave optimized for ions moving with velocity $\beta$=v/c=0.08 exhibits an order-of-magnitude reduction in radiofrequency (RF) losses into helium at 4.4 K compared to a superconducting niobium (Nb) cavity at the same frequency and temperature. Experimentally measured fields are among the highest to date for any Nb$_3$Sn-coated cavity, reaching a peak surface magnetic field of 105 mT. We also present a practical solution to the problem of cavity frequency tuning. Tuning by mechanical deformation has been a challenge with Nb$_3$Sn due to its brittle nature, however, using a set of techniques tailored to the properties of thin-film Nb$_3$Sn on Nb, we can repeatably tune the cavity to the ATLAS master clock frequency after it is cooled, while maintaining the excellent performance characteristics. The same Nb$_3$Sn cavity technology offers broad benefits for future ion accelerators.

## I. INTRODUCTION

Superconducting cavities are the leading technology for a large class of particle accelerators requiring proton and ion beams with energies from 5 MeV/u to 2.5 GeV/u or higher[1–5]. Their distinguishing attribute is their ability to support high accelerating gradients in continuous wave operation. Other beneficial properties include large transverse apertures and broad particle velocity acceptance. Well-developed ion linac cavity types include quarter-wave, half-wave and elliptical shapes, with the optimal type determined primarily by the ion velocity[6]. Cavities are arranged in fixed linear arrays, and the short ($\sim$ 1 m long) independently controlled cavities permit the acceleration of ions with variable energy over a wide range of mass-to-charge (*e.g.* $1 \leq A/q \leq 7$). This is essential for accelerators, such as ATLAS, that must deliver different ion species at variable energies.

We are developing a 145 MHz Nb$_3$Sn-coated quarter-wave cavity as a path to overcoming the performance limits of existing ATLAS Nb cavities. Nb$_3$Sn has roughly twice the critical temperature of Nb ($T_c$ = 18 K vs 9.2 K) and nearly twice the superheating field ($H_{SH} \approx$ 400 mT vs 240 mT)[7,8]. Because the BCS portion of the cavity surface resistance, $R_{BCS}$, decreases exponentially with temperature for $T \lesssim T_c/2$[9–11], the higher $T_c$ for Nb$_3$Sn makes it possible to achieve more than an order of magnitude reduction in RF losses with Nb$_3$Sn at 4.4 K. $H_{SH}$ is thought to present the fundamental limit on the highest sustainable magnetic field for SRF cavities, meaning that Nb$_3$Sn also offers the possibility for higher accelerating gradients.

---

a)T. Petersen and G. Eremeev contributed equally to this work.
b)University of California, Davis, Davis, United States of America

ATLAS beam physics requirements, set by the 49 operating Nb cavities, confine us to 50-150 MHz. However, we expect that the optimal frequencies for future 'green-field' Nb$_3$Sn-based ion linacs will be higher (*e.g.* 500 MHz), steming from the expected $R_{BCS}$ frequency and temperature dependence for Nb$_3$Sn. This would greatly reduce quarter- and half-wave (TEM-mode) cavity size and cost. At the same time, the reduction in RF power losses into helium at 4.4 K would enable the use of small, plug-in cryocoolers in place of large central liquid helium cryoplants and their complex distribution systems.[12,13] This prospect has motivated Nb$_3$Sn development for elliptical cavities for particles with $\beta \approx 1$[14–16].

## II. EXPERIMENTAL APPROACH

Electromagnetic (EM) simulations of the 145 MHz quarter-wave cavity RF volume and access ports were performed using CST Microwave Studio. Results for electric and magnetic fields and primary EM design parameters for the accelerating mode are in Fig. 1. and Table 1. Arrow sizes and directions in Fig. 1 indicate the relative strength and direction of the fields. The cavity shape includes the quarter-wave beam steering correction[17] and is generally similar to that for lower frequency niobium cavities operating in ATLAS.[18] Mechanical design and stress analyses were performed using COMSOL and AutoDesk Inventor.

The physical Nb$_3$Sn-coated cavity was formed starting from a Nb cavity fabricated by RadiaBeam and Argonne from high-purity (residual resistivity ratio, RRR>250) 3 mm-thick Nb sheet using well-developed techniques including die hydroforming, CNC machining, wire electron discharge machining and electron beam welding. Re-enforcing plates on the top and bottom of the cavity and flanges on the access ports were from lower-purity Nb (RRR$\approx$50) and a Nb (55%)-titanium al-loy, respectively. The interior surface of the Nb cavity was prepared for coating by removing 120 microns of niobium by electropolishing, similar for Nb cavities. Ultrasonic and high-pressure water cleaning were used to remove chemical residues.

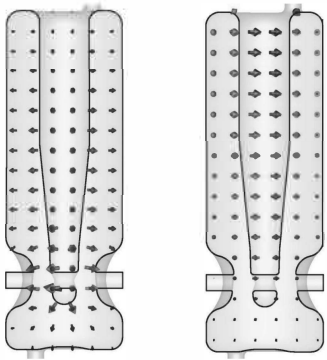

FIG. 1. Cavity section views showing electric (left) and magnetic (right) fields from CST MWS

TABLE I. Primary cavity electromagnetic and mechanical properties

| Parameter | Value | Unit |
|---|---|---|
| Frequency | 145.492 | MHz |
| Peak beta | 0.08 | |
| R/Q | 433 | Ohms |
| G | 29.5 | Ohms |
| Beam aperture | 3 | cm |
| Diameter | 21.6 | cm |
| Height | 63 | cm |
| Design voltage | 1.33 | MV |
| At the design voltage | | |
| Stored energy | 4.5 | Joules |
| $E_{PEAK}$ | 51.6 | MV/m |
| $B_{PEAK}$ | 60 | mT |

A ∼2 micron-thick Nb₃Sn layer was formed on the cavity RF surface at Fermilab by vapor diffusion in a high-vacuum furnace at 1100 °C. Procedures to minimize contamination of the Nb₃Sn layer during coating—particularly from NbTi flanges—follow established best practices for elliptical-cell cavities[8,19]. In the absence of prior low-beta experience with Nb₃Sn coatings, we adopted the elliptical-cell protocol with geometry-specific adaptations. The cavity was mounted with the long axis horizontal in the vacuum furnace, as in Fig.2 (left image), with a two tin/tin-chloride (Sn/SnCl₂) sources fixed to two of the six cavity ports. The furnace containing the cavity, tin sources and heater assemblies, was ramped to 1100 °C while the wire-heaters elevated the source temperatures to 1300 °C. The latter increases the tin vapor pressure inside the cavity helps promote stoichiometric A15 growth. The coated cavity was cooled under vacuum to 50 °C before venting with dry nitrogen and removal from the furnace.

Final cleaning of the coated Nb₃Sn cavity was similar to that for niobium cavities. High-pressure water rinsing with ultrapure water, clean room assembly, and a slow controlled pump down of the interior volume were used to minimize particulate contamination. During clean assembly the cavity was fitted with fixed RF drive and pickup probes with $Q_{ext}$ values of $9.5 \times 10^9$ and $2.9 \times 10^{11}$ respectively. Performance of the cavity at 4.4 K was measured before and after coating, to ensure that coating was on top of niobium of known good performance. All cold testing was performed by fully immersing the sealed and vertically oriented cavity into a bath of liquid helium contained inside a 60 cm diameter vacuum insulated and liquid nitrogen shielded dewar filled to a level of ∼ 90 cm. The helium bath pressure was maintained in the range of $1000 \pm 50$ Torr, with an associated helium bath temperature of $4.45 \pm 0.05$ K.

Cavity performance was measured as described in[10]. Continuous-wave RF power from a few μW to 17.5 W (transmitted) was delivered via the drive probe, with a phase-locked loop maintaining constant field. The power coupled into the cavity was taken as $P_{in} = P_{fwd} - P_{refl}$. The cavity $Q_0$ and $E_{acc}$ were then obtained from the measured input and pickup couplings, the RF decay time $\tau$, and the CST MWS–simulated relation between stored energy $U$ and accelerating gradient $E_{acc}$.

## III. FIELD PERFORMANCE RESULTS

The superconducting transition temperature was $T_c = 17.7 \pm 0.1$ K, determined primarily from VNA measurements of the cavity $S_{21}$ resonance at low forward power; temperature was read from calibrated silicon-diode thermometers distributed from top to bottom of the cavity. Magnetic flux expulsion—evidenced by an increase in the local $B$ field measured by three Bartington fluxgate magnetometers on the cavity exterior—occurred at the same temperature.

Fig. 3 compares 4.4 K performance before and after coating. The lowest curve is for the uncoated Nb cavity. The first test of the Nb₃Sn-coated cavity (middle curve) exhibited a high cavity quality factor at low fields, with $Q > 10^{10}$, a high maximum accelerating gradient, $E_{acc} = 12.5$ MV/m, ($B_{PEAK} = 95$ mT), but also, a pronounced downward slope in the qual-ity factor with increasing gradient. The relative improvement in cavity $Q$ after Nb$_3$Sn coating compared to the Nb cavity was one order of magnitude at very low fields; however, this decreased to a factor of 3 for $E_{acc} \geq 1$ MV/m.

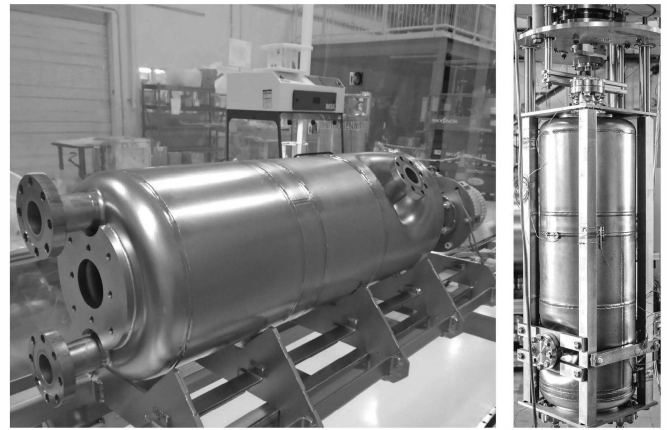

FIG. 2. Cavity with Sn source assembly after furnace coating (left) and cavity with Ti frame (right)

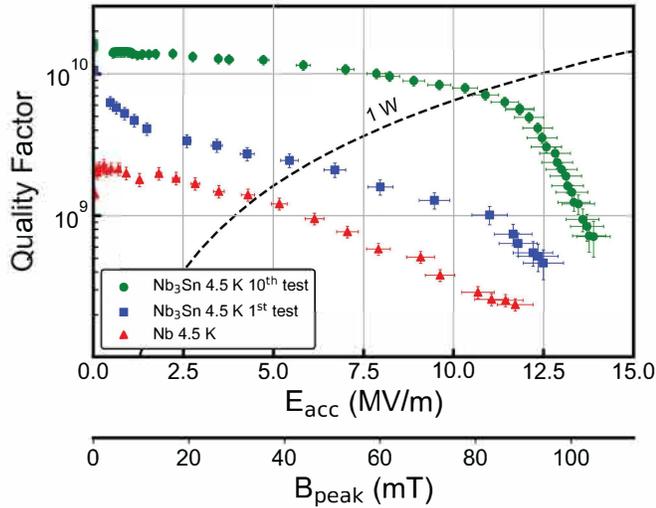

FIG. 3. Cavity Q versus accelerating gradient and peak magnetic field. 1 Watt constant transmitted power line shown for reference.

"Q-slope" can arise from several mechanisms, notably electron field emission (FE) and magnetic flux trapping[20–22]. We rule out a substantial contribution from FE based on the lack of detectable x-rays at the GM detector placed 40 cm from the cavity wall the outside of the cryostat . Maximum ambient (residual Earth's) magnetic fields over the volume of the cavity were 10 mG, as measured by three fluxgate magnetometers mounted inside the helium cryostat and against the cavity outer wall. In the worst case of 100% flux trapping with $B_{ambient}$=10 mG and using $R_s = 0.2$ n$\Omega$/mG[23,24] the added residual resistance would be 2 n$\Omega$. With geometry factor $G = Q_0 R_s = 29.5$ $\Omega$, this implies $Q_0 \gtrsim G/R_s \approx 1.4 \times 10^{10}$, i.e., substaintally higher than measured in the test.

Magnetic fields from Seebeck-driven thermocurrents near $T_c$ can degrade $Q_0$ of Nb$_3$Sn cavities. In a 218 MHz Nb$_3$Sn quarter-wave cavity, cooling through $T_c$ at a rate of 1 mK/s or slower with a maximum temperature gradient $\Delta T \sim 200$ mK minimized the $Q_0$ degradation due to thermocurrents. For the initial 145 MHz assembly, multiple warm/cool cycles through $T_c$ produced only minor changes in the $Q_0$ versus $E_{ACC}$ relative to the large overall Q-slope. These results suggest that the Q-slope is not primarily caused by residual Earth's field or thermocurrents.

After warming to room temperature, inspection of the testing setup revealed a magnetized RF cable along the cavity exterior, with magnetic field up to 350 mG. The high magnetic field at the cavity wall is linked to the Q-slope and underscores the importance of "magnetic hygiene" for Nb$_3$Sn cavities. Following the first test, all potentially ferrous materials in the test cryostat were verified to have fields at the surface $\leq 10$ mG.

After cable replacement, 4.4 K tests showed high $Q_0$ over a broad $E_{ACC}$ range. The low-field was $Q > 10^{10}$ with only a small slope up to $E_{ACC} \approx 10$ MV/m. We reached $E_{ACC} = 13.9$ MV/m ($B_{PEAK} = 105$ mT). The Q-slope above $E_{ACC} \approx 11$ MV/m is not field-emission–driven, based on a maximum x-ray measurement of 3 mR/h at 40 cm from the cavity wall. Notably, at the ATLAS design gradient, $E_{ACC} = 7.9$ MV/m, power dissipated into 4.4 K liquid helium is only 400 mW.

## IV. FREQUENCY TUNING

Cavities in ATLAS and in any accelerator with phase-locked RF systems must be tunable. For Nb cavities, this is a multistep process. Cavities are coarse-tuned during fabrication by plastic deformation (stretching or squeezing) to bring the frequency within the range of a slow tuner. Slow tuners are usually mechanical, integrated into the cavity and cryomodule, and adjust the cavity frequency during (4.4 K) operation. A Nb cavity that is out of the slow tuner's range can be warmed to room temperature and coarse-tuned with no practical impact on field performance. Unlike niobium, Nb$_3$Sn-coated SRF cavities have been shown to degrade due to mechanical tuning at room temperature, but may withstand larger tuning ranges at liquid helium temperatures[29,30].

Brittle Nb$_3$Sn coatings complicate tuning because any plastic deformation must be completed before coating. The final plastic tuning must be precise enough to accommodate the unavoidable frequency shifts caused by the coating process, evacuation of the interior volume, pressurization of the exterior helium volume after installation in the cryostat, and cooldown to 4.4 K. At 4.4 K, the slow tuner must provide sufficient range to bring the cavity onto resonance at the master clock frequency while avoiding deformations large enough to damage the Nb$_3$Sn layer.

In our approach, we measure the 4.4 K frequency of the uncoated Nb cavity. At room temperature, we apply a controlled axial plastic deformation (squeeze or stretch along the beamport axis) to tune toward the master frequency, guided by predicted frequency shifts for the steps above. After coating, we remeasure the 4.4 K frequency. Finally, at room temperature, we adjust the Ti frame using materials with known integrated coefficients of thermal expansion to cancel the residual frequency error. This approach minimizes total strain after coating and exploits the increase in Nb yield strength at low temperature. To avoid coupling to the Ti frame, the active tuner is mounted on top of the cavity. Small dynamic frequency corrections are made by pulling up or pushing down on the quarter-wave center conductor. By contrast, beam-port-mounted tuners (e.g., ATLAS and FRIB) operate in compression only and therefore can only lower the frequency.

Post-coating the measured 4.4 K frequency was low by 70 kHz. Achieving the required 380 µm axial stretch within the available 20 cm width of the Ti frame required a negative-CTE insert. We used non-magnetic ALLVAR alloy 30 with a fractional change of $\Delta L/L = 0.0035 \pm 0.0005$ from room temperature to 77 K—similar in magnitude but opposite in sign to that of copper. Two ALLVAR rods in the Ti frame (15.25 cm long, 1.91 cm diameter) expand by $\approx 600$ µm, providing ample range. Small brass spacers in series with the ALLVAR trimmed the final frequency to within 2 kHz, well inside the tuner and cavity elastic range. Tuner force–frequency data are in Fig. 4. Critically, the $Q_0$ versus $E_{ACC}$ showed no degradation after tuning, as shown for the 10th and most recent test in Fig. 3. Tests two through nine were similar, with $Q_0$ differing by a few percent depending on the cooldown rate through $T_c$.

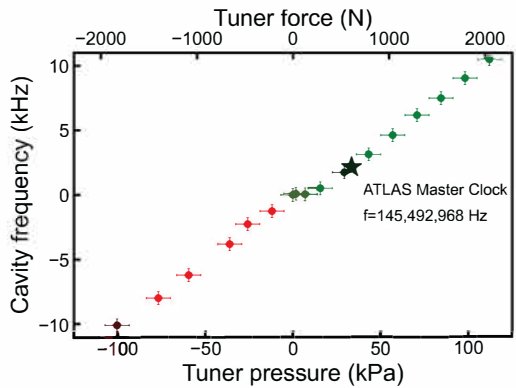

FIG. 4. Cavity frequency tuning using a pneumatic push-pull tuner

To understand reasons underlying the good 145 MHz cavity performance we review results for Nb witness samples placed at the port openings on opposite ends of the cavity and coated along with the cavity[25]. Scanning electron microscope (SEM) images of the samples show a uniform $\sim 1.5\,\mu$m thick layer of Nb$_3$Sn with uniform grain size of $\sim 0.8\,\mu$m. By contrast, our lower-performing 218 MHz cavity had greater end-to-end variability in grain size ($\sim 0.6$–$1.6\,\mu$m) and thickness ($\sim 1.5$–$2.7\,\mu$m). The reasons for the film-property differences remain unclear; two new 145 MHz cavities planned for coating may help clarify these effects.

Multipacting—though non-fundamental—has been unexpectedly persistent in the coated 145 MHz cavity: a low-level barrier at $E_{ACC} \approx 10$ kV/m has not yet been conditioned. We therefore use relatively high RF power to quickly push through the low-field barrier and measure $Q$ versus $E_{ACC}$ at higher fields. The fixed drive-probe coupling ($Q_{\text{ext}} = 9.5 \times 10^9$) may be insufficient and future Nb$_3$Sn tests will employ an adjustable power coupler.

## V. CONCLUSION

The peak magnetic field of 105 mT in the Nb$_3$Sn-coated 145 MHz cavity is among the highest fields reached in any Nb$_3$Sn cavity to date. The low dissipation at practical gradients together with the practical tuning system presented here demonstrate the substaintal benefit of Nb$_3$Sn-coated low-beta SRF cavities over niobium. We plan a phased deployment of Nb$_3$Sn cavities in ATLAS and other ion accelerators. As a first step, the 145 MHz cavity will serve as a rebuncher at the end of the ATLAS SRF linac to deliver beams to users; successful operation would motivate a multicavity accelerating cryomodule to replace the three remaining split-ring cryomodules. Beyond ATLAS, Nb$_3$Sn cavities could offer significant advantages over Nb for dedicated isotope-production facilities using ion and electron linacs[26,27] and for large proton- or light-ion SRF linacs for energy production or nuclear-waste transmutation[28].

## ACKNOWLEDGMENTS

This work is supported with resources of ANL's ATLAS facility, an Office of Science User Faciluty and by the U.S. Department of Energy, Office of Nuclear Physics, under Contract No. DE-AC02-06CH11357. This manuscript has been authored by FermiForward Discovery Group, LLC under Contract No. 89243024CSC000002 with the U.S. Department of Energy, Office of Science, Office of High Energy Physics.

## DATA AVAILABILITY STATEMENT

The data that support the findings of this study are available from the corresponding author upon reasonable request.